\documentclass{article}%
\usepackage{amsmath}
\usepackage{graphicx,epsfig,amsmath}
\usepackage{graphicx}%
\usepackage{amsfonts}%
\usepackage{amssymb}
\begin{document}

\title{{\Large Angular dependence of synchrotron radiation intensity}}
\author{V.G. Bagrov\thanks{Tomsk State University and Tomsk Institute of High Current
Electronics, Russia.}, V.G. Bulenok\thanks{Tomsk State Pedagogical University,
Russia.}, D.M. Gitman\thanks{Instituto de F\'{\i}sica, USP, Brasil, e-mail:
gitman@fma.if.usp.br .},
\and V.B. Tlyachev\thanks{Tomsk Institute of High Current Electronics, Russia.},
J.A. Jara\thanks{IFT, UNESP, Brasil.}, and A.T. Jarovoi\thanks{Tomsk State
Pedagogical University, Russia.}}
\date{\today            }
\maketitle

\begin{abstract}
The detailed analysis of angular dependence of the synchrotron radiation (SR)
is presented. In particular, we analyze the angular dependence of the integral
SR-intensity and peculiarities of the angular dependence of the first
harmonics SR. Studying spectral SR-intensities, we have discovered their
unexpected angular behavior, completely different from that of the integral
SR-intensity. Namely, for any given synchrotron frequency, maxima of the
spectral SR-intensities recede from the orbit plane with increasing particle
energy. Thus, in contrast with the integral SR-intensity, the spectral ones
have the tendency to deconcentrate themselves on the orbit plane.

\end{abstract}

%\textbf{Keywords:} synchrotron radiation, angular distribution \textbf{%
%Classification codes:}41.60. Ap

\section{Introduction}

At present the theory of synchrotron radiation (SR) is well developed and its
predictions are in good agreement with experiment \cite{1,2,3,4}. We recall
that the SR is created by charged particles, which are moving with velocities
$\upsilon$ along circles of radius $R$ in an uniform magnetic field $H$,
\begin{equation}
R=\frac{\beta E}{eH}=\frac{m_{0}c^{2}}{eH}\sqrt{\gamma^{2}-1}\,,\;\beta
=\frac{\upsilon}{c}\,,\;\;\gamma=(1-\beta^{2})^{-1/2}=\frac{E}{m_{0}c^{2}}%
\gg1\,. \label{1}%
\end{equation}
Here $E$ is the particle energy, $e$ is the charge, and $m_{0}$ the rest mass.
The radiation frequencies $\omega_{\nu}=\nu\omega_{0}\,,\;\nu=1,2,...,$ are
multiples of the synchrotron frequency $\omega_{0}=ceH/\,E\;. $ The spectral
SR-intensity (SR-intencity for a fixed radiation frequency) has maximum for
harmonics with $\nu\sim\gamma^{3}$. Two limiting cases, the non-relativistic
($\beta\ll1,\;E\simeq m_{0}c^{2}$) and\ the relativistic limits ($\beta
\sim1,\;E\gg m_{0}c^{2}$), are of particular interest.\ In the
non-relativistic case, only the first harmonic $\omega_{1}=\omega$ is
effectively emitted. The SR-intensity has a maximum in the direction of the
magnetic field. In the relativistic case, the integral SR-intensity (spectral
SR-intensity summed over the spectrum) is concentrated in the orbit plane
within a small angle $\Delta\theta\sim1/$\ $\gamma\ll1$\ . Thus, as the
electron energy increases, the integral SR-intensity tends to be concentrated
in the orbit plane. Any polarization component of the integral SR-intensity
has the same behavior. These results were first derived in the framework of
classical theory. Consideration in the framework of quantum theory does not
change essentially results of the classical analysis, since quantum
corrections are small \cite{1,2,3,4}.

However, one ought to say that the analysis of angular dependence of the
spectral and the integral SR-intensities was not done before in detail.
Recently this work was done by us, and in the present article we present
results of such an analysis. In Sect. II we analyze in detail angular
dependence of the integral SR-intensity. In Sect. III. we study peculiarities
of the angular dependence of the first harmonics SR. Studying spectral SR-
intensities (see Sect.IV), we have discovered their unexpected angular
behavior, completely different from that of the integral SR-intensity. Namely,
one can see that for any given synchrotron frequency, maxima of the spectral
SR-intensities recede from the orbit plane with increasing particle energy.
There exist limiting angles (at $\beta\rightarrow1$) for the maxima, which
depend on the synchrotron frequency. Thus, in contrast with the integral
SR-intensity, the spectral ones have the tendency to deconcentrate themselves
on the orbit plane. The analysis is done in the framework of classical theory,
but as was already mentioned above, quantum corrections cannot change the
results essentially.

\section{Angular dependence of integral SR- intensity}

In the SR theory one introduces polarization components $W_{i}$ ,\ $i=0,\pm
1,2,3$ of the integral SR-intensity \cite{1,2,3,4}. Here $W_{\pm1}$ are the
integral SR-intensities of the right ($+1$) and the left ($-1$) circular
polarization components respectively, whereas $W_{2}$ and $W_{3}$ are the so
called ''$\sigma$'' and ''$\pi$'' linear polarization components. The total
integral SR-intensity $W_{0}$ is defined as $W_{0}=W_{1}+W_{-1}=W_{\sigma
}+W_{\pi}\,.$ In the framework of the classical theory of SR one can find:%
\begin{align}
&  W_{i}=V_{0}\Phi_{i}(\beta),\ \;V_{0}=\frac{ce^{2}\beta^{4}}{R^{2}%
}=\frac{e^{4}H^{2}\beta^{2}(1-\beta^{2})}{m_{0}^{2}c^{3}}\,,\nonumber\\
&  \Phi_{i}(\beta)=\int_{0}^{\pi}F_{i}(\beta,\theta)\sin\theta d\theta
,\ F_{i}(\beta,\theta)=\sum_{\nu=1}^{\infty}f_{i}(\nu,\beta;\theta
)\,,\nonumber\\
&  f_{0}(\nu,\beta;\theta)=f_{-1}(\nu,\beta;\theta)+f_{1}(\nu,\beta
;\theta)=f_{2}(\nu,\beta;\theta)+f_{3}(\nu,\beta;\theta)\,. \label{2}%
\end{align}
Here $\theta$ is the angle between the $z$-axis and the radiation direction.
The sum over $\nu$ is just the sum over the spectrum, such that the
expressions inside the sum represent spectral distributions. The functions
$f_{i}(\nu,\beta;\theta)$ have the form:
\begin{align}
&  f_{\mp1}(\nu,\beta;\theta)=\frac{\nu^{2}}{2}\left[  J_{\nu}^{\prime}%
(z)\mp\frac{\cos\theta}{\beta\sin\theta}J_{\nu}(z)\right]  ^{2},\ \;z=\nu
\beta\sin\theta\,,\nonumber\\
&  f_{2}(\nu,\beta;\theta)=\nu^{2}J{_{\nu}^{\prime}}^{2}(z),\ \;f_{3}%
(\nu,\beta;\theta)=\frac{\nu^{2}\cos^{2}\theta}{\beta^{2}\sin^{2}\theta}%
J_{\nu}^{2}(z)\,. \label{3}%
\end{align}
Here $J_{\nu}(x)$ are Bessel functions of integer indices. The following
simple properties hold true:
\begin{equation}
f_{k}(\nu,\beta;\theta)=f_{k}(\nu,\beta;\pi-\theta),\ \;k=0,2,3;\;\ f_{-1}%
(\nu,\beta;\theta)=f_{1}(\nu,\beta;\pi-\theta)\,. \label{4a}%
\end{equation}
Thus, it is enough to study the functions $f_{k}(\nu,\beta;\theta),$
$k=0,2,3,$ at the interval $0\leq\theta\leq\pi/2$\ only, and between the
functions $f_{\pm1}$\ it is enough to study $f_{1}$ only.

Exact analytic expressions for the functions $F_{k}(\beta,\theta
)$\ ,$\;k=0,2,3$ were already known ~\cite{1,2,3,4}:
\begin{align}
&  F_{2}(\beta,\theta)=\frac{7-3\varepsilon}{16\varepsilon^{5/2}%
},\ \;\varepsilon=1-\beta^{2}\sin^{2}\theta,\ \;\frac{1}{\gamma^{2}}%
\leq\varepsilon<1\,,\nonumber\\
&  F_{3}(\beta,\theta)=\frac{(\gamma^{2}\varepsilon-1)(5-\varepsilon
)}{16(\gamma^{2}-1)\varepsilon^{7/2}},\ \;F_{0}(\beta,\theta)=\frac{(3-4\gamma
^{2})\varepsilon^{2}+6(2\gamma^{2}-1)\varepsilon-5}{16(\gamma^{2}%
-1)\varepsilon^{7/2}}\;. \label{5a}%
\end{align}
Expressions for the functions $F_{\pm1}$ can be found in the form:
\begin{equation}
F_{\pm1}(\beta,\theta)=\frac{1}{2}F_{0}(\beta,\theta)\pm\Psi(\beta\sin
\theta)\cos\theta\,,\;\;\Psi(x)=\frac{1}{2x}\frac{d}{dx}\sum_{\nu=1}^{\infty
}\nu J_{\nu}^{2}(\nu x)\,. \label{6a}%
\end{equation}

One can find that for any fixed $\beta$ all the functions $F_{i}(\beta
,\theta)$ have an extremum at $\theta=0\,.$ Moreover, the extremal values of
these functions do not depend on $\beta,$%
\begin{equation}
F_{-1}(\beta,0)=0,\ 2F_{0}(\beta,0)=2F_{1}(\beta,0)=4F_{2}(\beta
,0)=4F_{3}(\beta,0)=1\,.\label{7a}%
\end{equation}
The point $\theta=\pi/2$ provides an extremum for the functions $F_{k}%
$\ ,$\;k=0,2,3$ only. Here we have:
\begin{equation}
F_{0}(\beta,\pi/2)=F_{2}(\beta,\pi/2)=2F_{\pm}(\beta,\pi/2)=\frac{1}{16}%
\gamma^{3}(7\gamma^{2}-3),\;F_{3}(\beta,\pi/2)=0\,.\label{8a}%
\end{equation}
Therefore, for $F_{3}$ the point $\theta=\pi/2$ is an absolute minimum. For
any fixed $\beta$ the function $F_{2}(\beta,\theta)$ is a monotonically
increasing function of $\theta$ on the interval $0\leq\theta\leq\pi/2$ . Thus,
$\theta=0$ is an absolute minimum and $\theta=\pi/2$ is an absolute maximum of
this function. The maximum of the function $F_{2}$ increases as $E^{5}$ with
increasing particle energy $E.$

For $\gamma\leq\gamma_{0}^{(1)},\ (\beta\leq\beta_{0}^{(1)}),$%
\begin{equation}
\gamma_{0}^{(1)}=\sqrt{7/6}\approx1.0801,\;\ \beta_{0}^{(1)}=1/\sqrt{7}%
\approx0.378\,, \label{9a}%
\end{equation}
$F_{0}$ and $F_{1}$ are monotonically decreasing functions of $\theta$
($F_{0}$ on the interval $0\leq\theta\leq\pi/2$ and\ $F_{1}$ on the interval
$0\leq\theta\leq\pi$). Thus, at $\theta=0$ these functions have an absolute
maximum. The functions $F_{0}$ and $F_{1}$ have their absolute minima at
$\theta=\pi/2$ and $\theta=\pi$ respectively. Besides, $F_{1}(\beta,\pi)=0\,.$
For $\gamma_{0}^{(1)}<\gamma<\gamma_{0}^{(2)},\ (\beta_{0}^{(1)}<\beta
<\beta_{0}^{(2)}),$%
\begin{equation}
\gamma_{0}^{(2)}=\frac{\sqrt{3}+3\sqrt{2}}{5}\approx1.1949\,,\;\;\beta
_{0}^{(2)}=\sqrt{\frac{2}{3}(\sqrt{6}-2)}\approx0.5474\,, \label{10a}%
\end{equation}
the points $\theta=0,\ \pi/2$ are minima for $F_{0}$\thinspace, and the point
$\theta=\theta_{0}(\beta),$%
\begin{equation}
\sin^{2}\theta_{0}(\beta)=\frac{6\gamma^{2}(1-3\gamma^{2})+2\gamma^{2}%
\sqrt{15(15\gamma^{4}-22\gamma^{2}+9)}}{3(4\gamma^{2}-3)(\gamma^{2}%
-1)},\;\;0<\theta_{0}(\beta)<\pi/2\,, \label{11a}%
\end{equation}
provides a maximum for $F_{0}$ . For $\gamma_{0}^{(2)}<\gamma,\ (\beta
_{0}^{(2)}<\beta<1),$ the function $F_{0}$ has an absolute maximum at the
point $\theta=\pi/2$.

Denoting via $\theta_{0}^{(m)}(\beta)$ all the maximum points of $F_{0}%
$\thinspace, \ we may write:
\begin{equation}
\theta_{0}^{(m)}(\beta)=\left\{
\begin{array}
[c]{cl}%
0, & \beta\leq\beta_{0}^{(1)}\\
\theta_{0}(\beta), & \beta_{0}^{(1)}<\beta<\beta_{0}^{(2)}\\
\pi/2, & \beta_{0}^{(2)}\leq\beta<1
\end{array}
\right.  \,. \label{12a}%
\end{equation}
The plot of the function $\theta_{0}^{(m)}(\beta)$ see below:%
%TCIMACRO{\FRAME{ftbpFU}{3.0926in}{1.3889in}{0pt}{\Qcb{The function $\theta
%_{0}^{(m)}(\beta)$.}}{\Qlb{fig1}}{fig1r.eps}%
%width 3.0926in;  height 1.3889in;  depth 0pt;  original-width 2.9577in;
%original-height 1.3136in;  cropleft "0";  croptop "1";  cropright "1";
%cropbottom "0";  filename 'fig1r.eps';file-properties "XNPEU";}} }%
%BeginExpansion
\begin{figure}
[ptb]
\begin{center}
\includegraphics[
height=1.3889in,
width=3.0926in
]%
{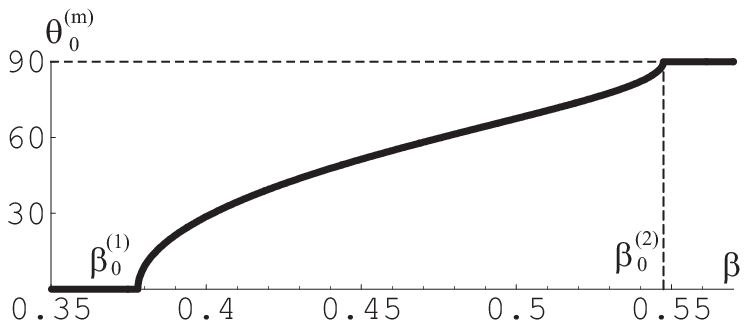}%
\caption{The function $\theta_{0}^{(m)}(\beta)$.}%
\label{fig1}%
\end{center}
\end{figure}
%EndExpansion

For any given $\beta\in(\beta_{0}^{(1)},1),$ the function $F_{1}$ has its
maximum at the point $\theta=\theta_{1}(\beta)$\ $,\;$ $0<\theta_{1}%
(\beta)<\pi/2$ . Denoting via $\theta_{1}^{(m)}(\beta)$ all the maximum points
of $F_{1}$\thinspace, we may write:
\begin{equation}
\theta_{1}^{(m)}(\beta)=\left\{
\begin{array}
[c]{cl}%
0, & \beta\leq\beta_{0}^{(1)}\\
\theta_{1}(\beta), & \beta_{0}^{(1)}<\beta<1
\end{array}
\right.  \,. \label{13a}%
\end{equation}
At the moment, there is no analytical expression for $\theta_{1}(\beta)$
similar to (\ref{11a}) for $\theta_{0}(\beta)$. However, one can see that the
function $\theta_{1}(\beta)$ is a monotonically increasing function of
$\beta\in\left[  \beta_{0}^{(1)},1\right]  .$ \thinspace For $\beta
\rightarrow1$\ there is an asymptotic form
\begin{equation}
\theta_{1}(\beta)\approx\pi/2-a_{1}/\gamma\,, \label{14a}%
\end{equation}
where $\alpha_{1}\approx0.2672$ is a root of the equation (see \cite{1})
\begin{equation}
5\pi a_{1}(5+12a_{1}^{2})\sqrt{3}+64(5a_{1}^{2}-1)\sqrt{1+a_{1}^{2}}=0\ .
\label{15a}%
\end{equation}
The plot of the function $\theta_{1}^{(m)}(\beta)$ see below:
%TCIMACRO{\FRAME{ftbpFU}{3.384in}{1.446in}{0pt}{\Qcb{The function $\theta
%_{1}^{(m)}(\beta)$}}{\Qlb{fig2}}{fig2r.eps}%
%width 3.384in;  height 1.446in;  depth 0pt;  original-width 3.2379in;
%original-height 1.3681in;  cropleft "0";  croptop "1";  cropright "1";
%cropbottom "0";  filename 'fig2r.eps';file-properties "XNPEU";}} }%
%BeginExpansion
\begin{figure}
[ptb]
\begin{center}
\includegraphics[
height=1.446in,
width=3.384in
]%
{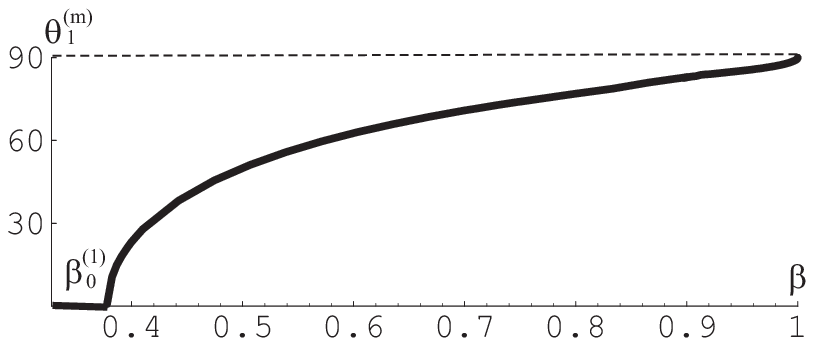}%
\caption{The function $\theta_{1}^{(m)}(\beta)$}%
\label{fig2}%
\end{center}
\end{figure}
%EndExpansion

For $\beta\leq\beta_{3}$ $(\gamma\leq\gamma_{3}),$
\begin{equation}
\beta_{3}=\frac{2}{\sqrt{15}}\approx0.5164\,,\ \gamma_{3}=\sqrt{\frac{15}{11}%
}\approx1.1678\ , \label{16a}%
\end{equation}
$F_{3}$ is a monotonically decreasing function on the interval $0\leq
\theta\leq\pi/2$ . The point $\theta=0$ provides the absolute maximum for this
function. For $1>\beta>\beta_{3}\,,\;(\gamma>\gamma_{3})\,,$\ the points
$\theta=0$ and $\theta=\theta_{3}(\beta)$ provide the minimum and the maximum
respectively for $F_{3}$ ,%
\begin{equation}
\sin^{2}\theta_{3}(\beta)=\frac{\sqrt{5(125\gamma^{4}-34\gamma^{2}%
+5)}-19\gamma^{2}-5}{6(\gamma^{2}-1)}\,,\ \;0<\theta_{3}(\beta)<\pi/2\,.
\label{17a}%
\end{equation}
Denoting via $\theta_{3}^{(m)}(\beta)$ all the maximum points of $F_{3}%
$\thinspace, \ we may write:
\begin{equation}
\theta_{3}^{(m)}(\beta)=\left\{
\begin{array}
[c]{cl}%
0, & \beta\leq\beta_{3}\\
\theta_{3}(\beta), & \beta_{3}<\beta<1
\end{array}
\right.  \,. \label{18a}%
\end{equation}
For $\beta\rightarrow1$ the following asymptotic expression holds true:%
\begin{equation}
\theta_{3}^{(m)}\approx\pi/2-\frac{1}{\gamma}\sqrt{\frac{2}{5}}. \label{19a}%
\end{equation}
The plot of the function $\theta_{3}^{(m)}(\beta)$ see below:
%TCIMACRO{\FRAME{ftbpFU}{3.2093in}{1.3759in}{0pt}{\Qcb{The function $\theta
%_{3}^{(m)}(\beta)$}}{\Qlb{fig3}}{fig3r.eps}%
%width 3.2093in;  height 1.3759in;  depth 0pt;  original-width 3.0701in;
%original-height 1.3007in;  cropleft "0";  croptop "1";  cropright "1";
%cropbottom "0";  filename 'fig3r.eps';file-properties "XNPEU";}}}%
%BeginExpansion
\begin{figure}
[ptb]
\begin{center}
\includegraphics[
height=1.3759in,
width=3.2093in
]%
{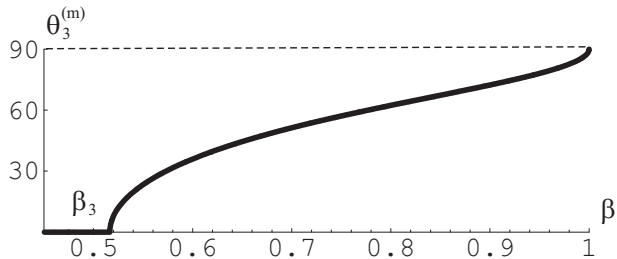}%
\caption{The function $\theta_{3}^{(m)}(\beta)$}%
\label{fig3}%
\end{center}
\end{figure}
%EndExpansion

\section{Angular dependence of spectral SR-intensity}

\subsection{First harmonic radiation}

The angular distribution of SR from the first harmonic ($\nu=1$) is distinctly
different from that of the higher harmonics ($\nu\geq2$) . Previously it was
known~\cite{1,2,3,4} that: a) The first harmonic alone contributes essentially
to the radiation in the directions $\theta=0,\pi\,. $ b) In the
nonrelativistic case ($\beta\sim0$), the radiation is maximal exactly in these directions.

Let us consider Eqs. (\ref{3}) for the first harmonic,%
\begin{align}
&  f_{\mp1}(1,\beta;\theta)=\frac{1}{2}\left[  J_{1}^{\prime}(z)\mp
\frac{\cos\theta}{x}J_{1}(z)\right]  ^{2},\ \;z=\beta\sin\theta\,,\nonumber\\
&  f_{2}(1,\beta;\theta)=J{_{1}^{\prime}}^{2}(z),\;\;f_{3}(1,\beta
;\theta)=\frac{\cos^{2}\theta}{z^{2}}J_{1}^{2}(z)\,. \label{20}%
\end{align}

In the nonrelativistic case ($\beta=0$) we get:
\begin{equation}
f_{\mp1}(1,0;\theta)=\frac{1}{8}(1\mp\cos\theta)^{2},\;\ f_{2}(1,0;\theta
)=\frac{1}{4},\;\ f_{3}(1,0;\theta)=\frac{\cos^{2}\theta}{4}\,. \label{21}%
\end{equation}
Thus, in this case, the radiation components $W_{0},\;W_{2},$ and $W_{3}$ peak
at $\theta=0,\pi,$ whereas $W_{\sigma}$ does not depend on $\theta$ at all.

Analyzing the expressions (\ref{20}), one can see that the functions
$f_{k}(1,\beta;\theta),\;k=0,1,2,3,$ peak at $\theta=0$ for any $\beta$
(including $\beta\rightarrow1$). Thus, the corresponding radiation components
$W_{k}$ are maximal at $\theta=0\,\ $for any $\beta\,.$

Besides, at any fixed $\theta\neq0,\pi,$ the functions $f_{k}(1,\beta
;\theta)\ ,\;k=-1,0,1,2,3,$ decrease monotonically with increasing $\beta$.
Thus, the radiation from the first harmonic has the tendency to line up in the
direction $\ \theta=0,\pi$ with increasing electron energy ($\beta
\rightarrow1$). This behavior of the first harmonic radiation is completely
opposite to that of the total SR-intensity in the ultrarelativistic case (as
was already said, see the previous Section, the latter radiation tends to be
concentrated in the orbit plane). All the functions (\ref{20}) have finite
limits as $\beta\rightarrow1$. See below the plot of the function (\ref{20})
at $\beta=0,1.$
%TCIMACRO{\FRAME{ftbpFU}{3.5993in}{2.1828in}{0pt}{\Qcb{$\overline{f}_{k}$--the
%functions $f_{k}(1,\beta;\theta)$ for $\beta=0$; $f_{k}$-- the functions
%$f_{k}(1,\beta;\theta)$ for $\beta=1$}}{\Qlb{fig4}}{fig4r.eps}%
%width 3.5993in;  height 2.1828in;  depth 0pt;  original-width 3.4463in;
%original-height 2.0799in;  cropleft "0";  croptop "1";  cropright "1";
%cropbottom "0";  filename 'fig4r.eps';file-properties "XNPEU";}}}%
%BeginExpansion
\begin{figure}
[ptb]
\begin{center}
\includegraphics[
height=2.1828in,
width=3.5993in
]%
{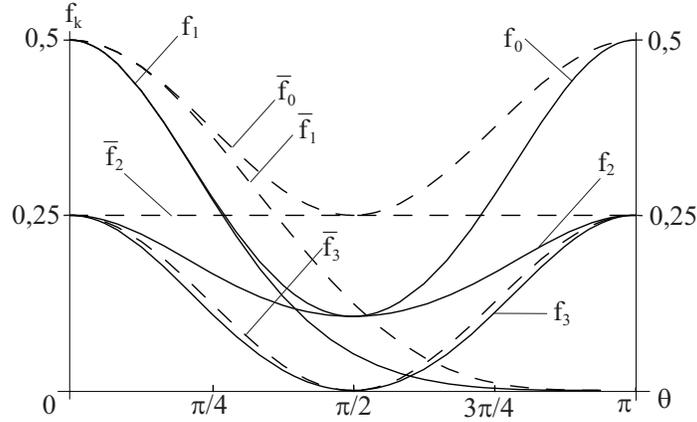}%
\caption{$\overline{f}_{k}$--the functions $f_{k}(1,\beta;\theta)$ for
$\beta=0$; $f_{k}$-- the functions $f_{k}(1,\beta;\theta)$ for $\beta=1$}%
\label{fig4}%
\end{center}
\end{figure}
%EndExpansion

\subsection{Higher harmonic radiation}

To study the angular dependence of higher harmonic ($\nu>1$) radiation, we
have to analyze the angular dependence of the functions $f_{k}(\nu
,\beta;\theta)$ for $\nu>1.$

First of all, one has to remark that all the functions $f_{k}(\nu,\beta
;\theta)$, $\nu>1$, vanish at $\theta=0,\pi$. Thus, they have the mentioned
absolute minimum at these points. By virtue of property (\ref{4a}) the
functions $f_{s}(\nu,\beta;\theta),\ s=0,2,3,$ have two symmetric maxima at
the points
\begin{equation}
\theta_{s}^{\nu}(\beta)=\pi/2\mp\delta_{s}(\nu,\beta),\ \;s=0,2,3\,.
\label{22}%
\end{equation}
The functions $f_{\pm1}(\nu,\beta;\theta)$ have maxima at the points
\begin{equation}
\theta_{\pm1}^{\nu}(\beta)=\pi/2\mp\delta_{1}(\nu,\beta)\,. \label{23}%
\end{equation}
All $\delta_{k}(\nu,\beta),\ k=0,1,2,3$ are non-decreasing functions of
$\beta$ for any given $\nu$\thinspace, and for any given $\beta$ they are
non-increasing functions of $\nu$ . At the same time,
\begin{equation}
0\leq\delta_{k}(\nu,\beta)<\pi/2,\ \;k=0,1,2,3\,, \label{24}%
\end{equation}
and
\begin{equation}
\lim_{\beta\rightarrow1}\delta_{k}(\nu,\beta)=\delta_{k}(\nu,1)=\delta
_{k}^{\nu}<\pi/2,\ \;k=0,1,2,3\,. \label{25}%
\end{equation}
The quantities $\delta_{k}^{\nu}$ are maxima for $\delta_{k}(\nu,\beta)$ at
fixed $k,\nu\,.$

Thus, for each harmonic ($\nu>1$) and for each polarization component, the
angular distribution of the SR-intensity has its own maximum. All these maxima
have the tendency to recede away from the orbit plane with increasing particle
energy. Therefore, as in the case $\nu=1$, the spectral SR-intensities for
$\nu>1$ have the tendency to deconcentration from the orbit plane with
increasing particle energy.

Below we study the functions $\delta_{k}(\nu,\beta)$ in detail.

Let $k=0,2$ . Then
\begin{equation}
\delta_{k}(\nu,\beta)=0,\ \;\beta<\beta_{k}^{\nu},\ \,k=0,2\,. \label{26}%
\end{equation}
Here $\beta_{0}^{\nu}$ and $\beta_{2}^{\nu}$ are respectively roots of the
transcendental equations%
\begin{equation}
2\,J_{\nu}(\nu\,\beta)=\beta\,\left[  \sqrt{\nu^{2}\,(1-\beta^{2})^{2}-4}%
+\nu\,(1-\beta^{2})\right]  \,J_{\nu}^{\prime}(\nu\,\beta)\,, \label{27}%
\end{equation}
and
\begin{equation}
\nu(1-\beta^{2})J_{\nu}(\nu\beta)-\beta J_{\nu}^{\prime}(\nu\beta)=0\,.
\label{28}%
\end{equation}
The condition $\beta\leq1$ implies that $\beta_{0}^{\nu}$ and $\beta_{2}^{\nu
}$ are unique. The following inequality holds true
\begin{equation}
\beta_{0}^{\nu}<\beta_{2}^{\nu}\,. \label{29}%
\end{equation}
For $\gamma_{k}^{\nu}=\left[  1-\left(  \beta_{k}^{\nu}\right)  ^{2}\right]
^{-1/2},$ one can find the asymptotic (at $\nu\gg1$) expressions:
\begin{align}
&  \gamma_{0}^{\nu}\approx\left(  \frac{\nu}{a_{0}}\right)  ^{1/3}%
,\ \;\gamma_{2}^{\nu}\approx\left(  \frac{\nu}{a_{2}}\right)  ^{2/3}%
\,,\nonumber\\
&  a_{0}=3z_{0}\approx0,7332,\ \;a_{2}=\left[  \frac{16\pi^{3}}{\Gamma
^{6}(1/3)\sqrt{3}}\right]  ^{1/4}\approx0,9382\,, \label{30a}%
\end{align}
Here $z_{0}$ is a root of the transcendental equation
\[
3z_{0}K_{2/3}(z_{0})-K_{1/3}(z_{0})=0,\ \;z_{0}\approx0,2444\,,
\]
where $K_{\mu}(x)$ are the Macdonald functions.

For $\beta>\beta_{0}^{\nu}$ , the function $\delta_{0}(\nu,\beta)$ is defined
as a solution of the transcendental equation
\begin{align}
&  2\,J_{\nu}(\nu\beta\cos\delta_{0})=\left\{  \nu\,[1-\beta^{2}+(1+\beta
^{2})\sin^{2}\delta_{0}]+\right. \nonumber\\
&  +\left.  \sqrt{\nu^{2}\,[1-\beta^{2}+(1+\beta^{2})\sin^{2}\delta_{0}%
]^{2}-4}\right\}  \,J_{\nu}^{\prime}(\nu\beta\cos\delta_{0})\,\beta
\,\cos{\delta_{0}\,}. \label{31a}%
\end{align}
(One can see that the equation (\ref{27}) is a particular case of (\ref{31a})
at $\delta_{0}=0$). Here $\delta_{0}(\nu,\beta)$ is a monotonically increasing
function of $\beta$ for each given $\nu$ . The maximum value $\delta_{0}^{\nu
}$ of $\delta_{0}(\nu,\beta)$ is a solution of the equation (\ref{31a}) for
$\beta=1$. There is the asymptotic (at $\nu\gg1$) expression
\begin{equation}
\delta_{0}^{\nu}\approx\left(  \frac{b_{0}}{\nu}\right)  ^{1/3},\ \ b_{0}%
=3p_{0}\approx0,3066\,, \label{32}%
\end{equation}
where $p_{0}$ is a root of the transcendental equation
\begin{equation}
6p_{0}K_{2/3}(p_{0})-K_{1/3}(p_{0})=0,\ \;p_{0}\approx0,1022\,. \label{33}%
\end{equation}

For $\beta>\beta_{2}^{\nu}\,,$ the function $\delta_{2}(\nu,\beta)$ has the
form:
\begin{equation}
\delta_{2}(\nu,\beta)=\arccos(\beta_{2}^{\nu}/\beta),\ \;\beta\in\left(
\beta_{2}^{\nu},1\right)  \,. \label{34}%
\end{equation}
Therefore, the maximum value $\delta_{2}^{\nu}$ of $\delta_{2}(\nu,\beta)$ at
the point $\beta=1$ is:
\begin{equation}
\delta_{2}^{\nu}=\delta_{2}(\nu,1)=\arccos\beta_{2}^{\nu}\,. \label{35}%
\end{equation}

According to (\ref{30a}), we have the asymptotic (at $\nu\gg1$) expression
\begin{equation}
\delta_{2}^{\nu}\approx1/\gamma_{2}^{\nu}\;. \label{36}%
\end{equation}

The functions $\delta_{1}(\nu,\beta)$ and $\delta_{3}(\nu,\beta)$ behave
similarly to $\delta_{0}(\nu,\beta)$ and $\delta_{2}(\nu,\beta)$. Namely,
$\delta_{1}(\nu,\beta)$ and $\delta_{3}(\nu,\beta)$ are defined as solutions
of the transcendental equations
\begin{align}
&  (\nu\sin\delta_{1}-1-\nu\beta^{2}\sin\delta_{1}\cos^{2}\delta_{1})J_{\nu
}(\nu\beta\cos\delta_{1})\nonumber\\
&  \,=J_{\nu}^{\prime}(\nu\beta\cos\delta_{1})\beta(1-\nu\sin\delta_{1}%
)\sin\delta_{1}\cos\delta_{1}\,, \label{37}%
\end{align}
and
\begin{equation}
J_{\nu}^{\prime}(\nu\beta\cos\delta_{3})\nu\beta\sin^{2}\delta_{3}\cos
\delta_{3}=J_{\nu}(\nu\beta\cos\delta_{3})\,, \label{38}%
\end{equation}
respectively. For each given $\nu,$ these functions are bounded and
monotonically increasing functions of $\beta\in\left[  0,1\right]  $ ,%
\begin{align}
&  \arcsin(1/\nu)=\delta_{1}(\nu,0)\leq\delta_{1}(\nu,\beta)\leq\delta
_{1}^{\nu}=\delta_{1}(\nu,1)\,,\nonumber\\
&  \arcsin(1/\sqrt{\nu})=\delta_{3}(\nu,0)\leq\delta_{3}(\nu,\beta)\leq
\delta_{3}^{\nu}=\delta_{3}(\nu,1)\,. \label{39}%
\end{align}
For each given $\beta,$ these functions decrease monotonically with increasing
$\nu$.

At $\nu\gg1$ we get the following asymptotic expressions:
\begin{equation}
\delta_{1}^{\nu}\approx\left(  \frac{b_{1}}{\nu}\right)  ^{1/3},\ \;b_{1}%
=3p_{1}\approx0.3933,\ \;\delta_{3}^{\nu}\approx\left(  \frac{a_{0}}{\nu
}\right)  ^{1/3}\;, \label{41}%
\end{equation}
where $a_{0}$ is defined by (\ref{30a}), and $p_{1}\approx0,1311.4$ \ is a
root of the transcendental equation
\begin{equation}
(3p_{1}-1)K_{1/3}(p_{1})+3p_{1}K_{2/3}(p_{1})=0\,. \label{42}%
\end{equation}

The following inequalities hold true:
\begin{equation}
\delta_{3}(\nu,\beta)>\delta_{1}(\nu,\beta)>\delta_{0}(\nu,\beta)\geq
\delta_{2}(\nu,\beta). \label{43}%
\end{equation}

The threshold values $\gamma_{0}^{\nu},$ $\gamma_{2}^{\nu}$ and the extremal
values $\delta_{\nu}^{k}$ (in Celsius) for some $\nu,$ are given on the Table
1:\begin{table}[ptb]%
\[%
\begin{array}
[c]{rrrrrrrrrrr}%
\nu & 2 & 3 & 4 & 5 & 6 & 7 & 10 & 15 & 20 & 25\\
\gamma_{0}^{\nu} & 1.00 & 1.22 & 1.40 & 1.54 & 1.67 & 1.79 & 2.08 & 2.46 &
2.75 & 3.00\\
\gamma_{2}^{\nu} & 1.59 & 2.10 & 2.55 & 2.97 & 3.36 & 3.73 & 4.75 & 6.25 &
7.59 & 8.82\\
\delta_{0}^{\nu} & 45.50 & 36.22 & 31.29 & 28.11 & 25.84 & 24.10 & 20.66 &
17.48 & 15.59 & 14.30\\
\delta_{1}^{\nu} & 45.88 & 36.83 & 32.02 & 28.91 & 26.68 & 24.98 & 21.57 &
18.39 & 16.48 & 15.16\\
\delta_{2}^{\nu} & 38.84 & 28.44 & 23.06 & 19.67 & 17.30 & 15.54 & 12.14 &
9.20 & 7.57 & 6.51\\
\delta_{3}^{\nu} & 49.83 & 41.09 & 36.29 & 33.11 & 30.80 & 29.00 & 25.34 &
21.83 & 19.69 & 18.19\\
\nu & 30 & 35 & 40 & 45 & 50 & 100 & 200 & 300 & 400 & 500\\
\gamma_{0}^{\nu} & 3.21 & 3.40 & 3.58 & 3.74 & 3.88 & 4.98 & 6.35 & 7.31 &
8.07 & 8.70\\
\gamma_{2}^{\nu} & 9.98 & 11.07 & 12.10 & 13.10 & 14.06 & 22.38 & 35.58 &
46.66 & 56.54 & 65.63\\
\delta_{0}^{\nu} & 12.34 & 12.59 & 11.98 & 11.47 & 11.03 & 8.60 & 6.74 &
5.86 & 5.31 & 4.92\\
\delta_{1}^{\nu} & 14.18 & 13.41 & 12.77 & 12.24 & 11.79 & 9.24 & 7.28 &
6.34 & 5.75 & 5.33\\
\delta_{2}^{\nu} & 5.75 & 5.18 & 4.74 & 4.38 & 4.08 & 2.56 & 1.61 & 1.23 &
1.01 & 0.87\\
\delta_{3}^{\nu} & 17.06 & 16.16 & 15.43 & 14.81 & 14.28 & 11.26 & 8.90 &
7.76 & 7.04 & 6.53
\end{array}
\]
\caption{Threshold values $\gamma_{0}^{\nu}$, $\gamma_{2}^{\nu}$ and limit
values $\delta\nu^{k}$ (in degrees)}%
\end{table}

\newpage

The Fig. 5 presents the plots of $\delta_{k}(\nu,\beta)$ ($\delta_{k}$ are
given in Celsius).
%~\ref{fig6}
The Fig. 6 presents the plots of $f_{k}(\nu,\beta;\theta)$ at $\beta=1$ and at
$\nu=1-5,10$.
%%%%%%%%%%%%%%%%%%%%%%%%%%%%%%%%%%%%%%%%%%%%%%%%%%%%%%%%%%%%%%%%%%%%%%%%%%%%%
%TCIMACRO{\FRAME{dtbpFU}{2.9023in}{1.849in}{0pt}{\Qcb{}}{}{fig_d_0r.eps}%
%width 2.9023in;  height 1.849in;  depth 0pt;  original-width 3.7369in;
%original-height 2.3705in;  cropleft "0";  croptop "1";  cropright "1";
%cropbottom "0";  filename 'fig_d_0r.eps';file-properties "XNPEU";}}}%
%BeginExpansion
\begin{center}
\includegraphics[
height=1.849in,
width=2.9023in
]%
{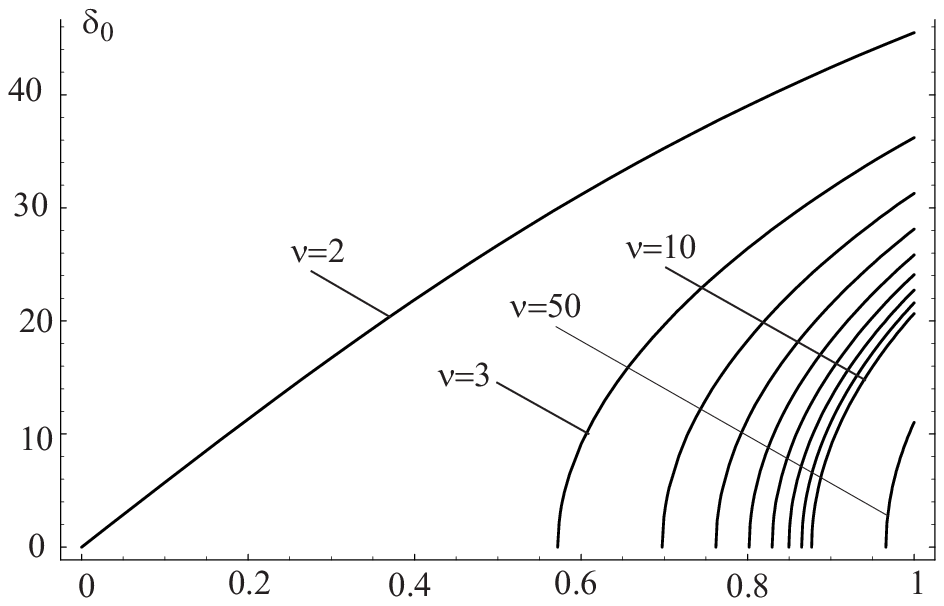}%
\end{center}
%EndExpansion%
%TCIMACRO{\FRAME{dtbpF}{2.8599in}{1.8464in}{0pt}{}{}{fig_d_1r.eps}%
%width 2.8599in;  height 1.8464in;  depth 0pt;  original-width 3.8726in;
%original-height 2.4915in;  cropleft "0";  croptop "1";  cropright "1";
%cropbottom "0";  filename 'fig_d_1r.eps';file-properties "XNPEU";}}}%
%BeginExpansion
\begin{center}
\includegraphics[
height=1.8464in,
width=2.8599in
]%
{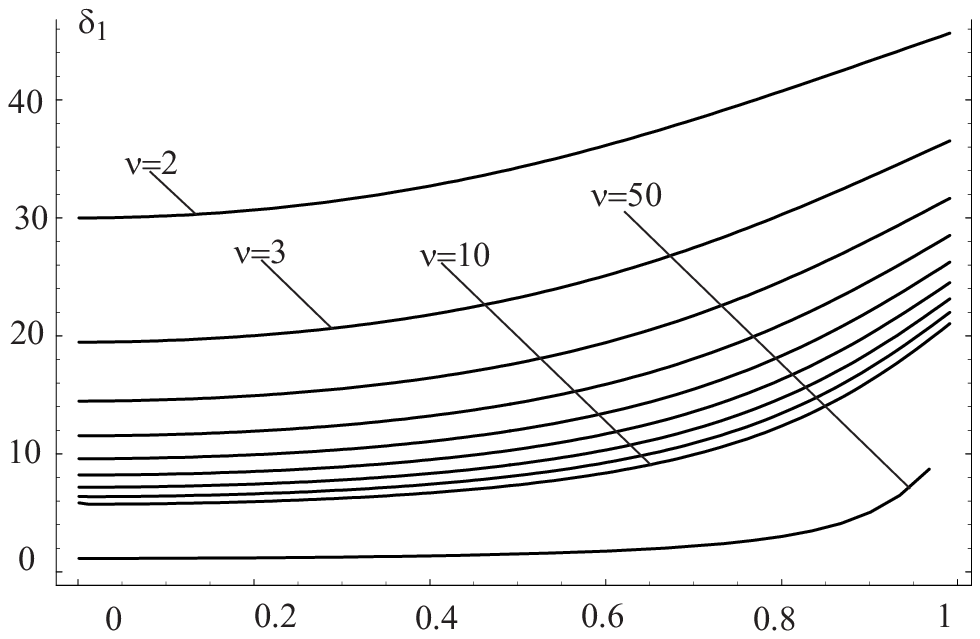}%
\end{center}
%EndExpansion%
%TCIMACRO{\FRAME{dtbpFU}{2.8331in}{1.8602in}{0pt}{\Qcb{}}{}{fig_d_2r.eps}%
%width 2.8331in;  height 1.8602in;  depth 0pt;  original-width 3.8303in;
%original-height 2.5062in;  cropleft "0";  croptop "1";  cropright "1";
%cropbottom "0";  filename 'fig_d_2r.eps';file-properties "XNPEU";}}}%
%BeginExpansion
\begin{center}
\includegraphics[
height=1.8602in,
width=2.8331in
]%
{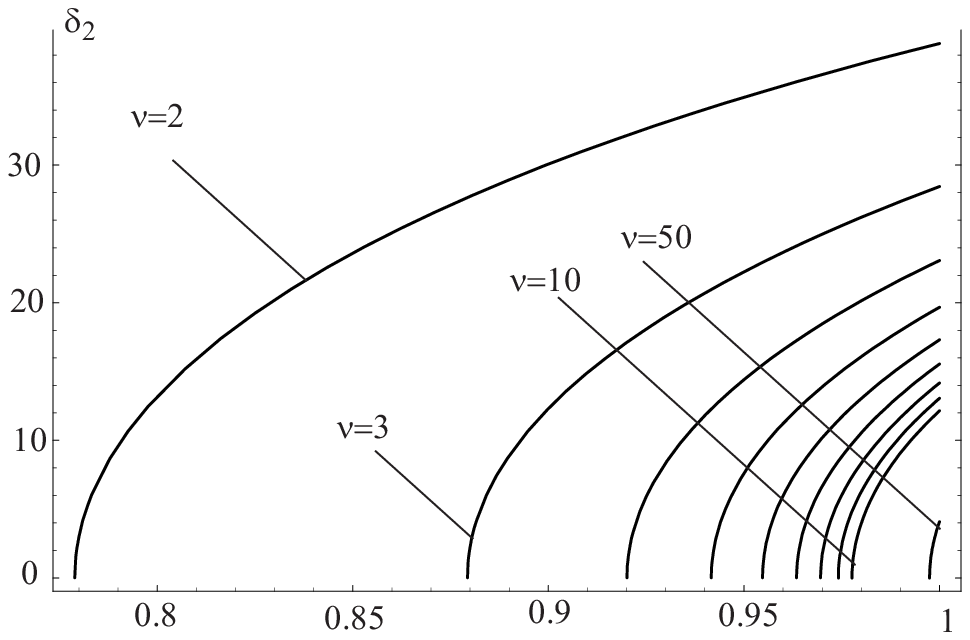}%
\end{center}
%EndExpansion%
%TCIMACRO{\FRAME{ftbphFU}{2.8418in}{1.8663in}{0pt}{\Qcb{The functions
%$\delta_{k}(\nu,\beta),(k=0,1,2,3)$ at $\nu=1-10,50$.}}{\Qlb{fig5}%
%}{fig_d_3r.eps}{\special{ language "Scientific Word";  type "GRAPHIC";
%width 2.8418in;  height 1.8663in;  depth 0pt;  original-width 3.8303in;
%original-height 2.5062in;  cropleft "0";  croptop "1";  cropright "1";
%cropbottom "0";  filename 'fig_d_3r.eps';file-properties "XNPEU";}}}%
%BeginExpansion
\begin{figure}
[ptbh]
\begin{center}
\includegraphics[
height=1.8663in,
width=2.8418in
]%
{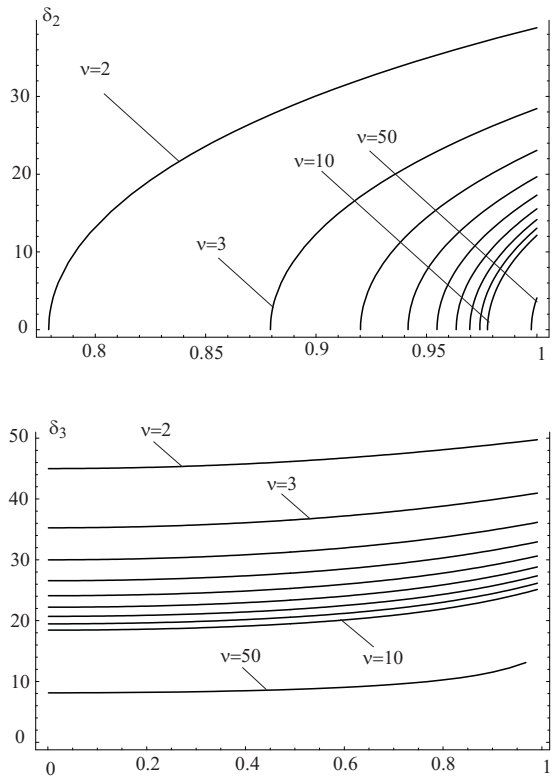}%
\caption{The functions $\delta_{k}(\nu,\beta),(k=0,1,2,3)$ at $\nu=1-10,50$.}%
\label{fig5}%
\end{center}
\end{figure}
%EndExpansion

\bigskip%

%TCIMACRO{\FRAME{dtbpF}{2.8478in}{1.8351in}{0pt}{}{}{fig_d0_1.eps}%
%width 2.8478in;  height 1.8351in;  depth 0pt;  original-width 2.7224in;
%original-height 1.7452in;  cropleft "0";  croptop "1";  cropright "1";
%cropbottom "0";  filename 'fig_d0_1.eps';file-properties "XNPEU";}}}%
%BeginExpansion
\begin{center}
\includegraphics[
height=1.8351in,
width=2.8478in
]%
{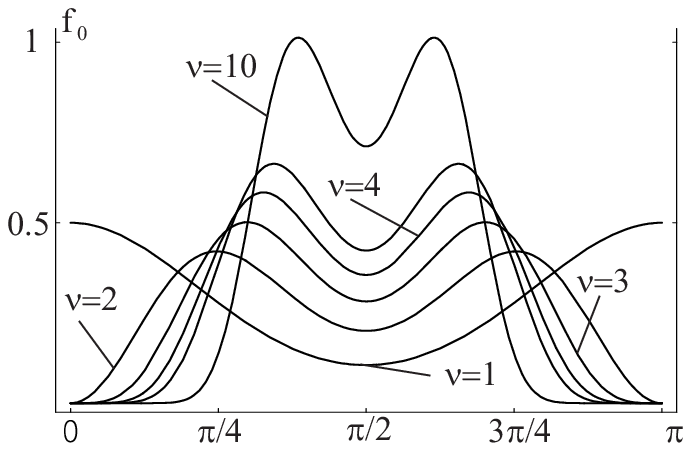}%
\end{center}
%EndExpansion%
%TCIMACRO{\FRAME{dtbpF}{2.8193in}{1.8049in}{0pt}{}{}{fig_d1_1.eps}%
%width 2.8193in;  height 1.8049in;  depth 0pt;  original-width 2.6948in;
%original-height 1.7141in;  cropleft "0";  croptop "1";  cropright "1";
%cropbottom "0";  filename 'fig_d1_1.eps';file-properties "XNPEU";}}}%
%BeginExpansion
\begin{center}
\includegraphics[
height=1.8049in,
width=2.8193in
]%
{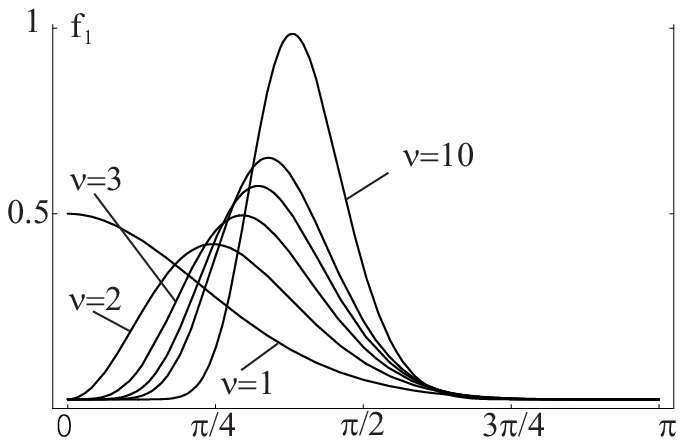}%
\end{center}
%EndExpansion%
%TCIMACRO{\FRAME{dtbpF}{2.8349in}{1.7789in}{0pt}{}{}{fig_d2_1.eps}%
%width 2.8349in;  height 1.7789in;  depth 0pt;  original-width 2.7095in;
%original-height 1.689in;  cropleft "0";  croptop "1";  cropright "1";
%cropbottom "0";  filename 'fig_d2_1.eps';file-properties "XNPEU";}}}%
%BeginExpansion
\begin{center}
\includegraphics[
height=1.7789in,
width=2.8349in
]%
{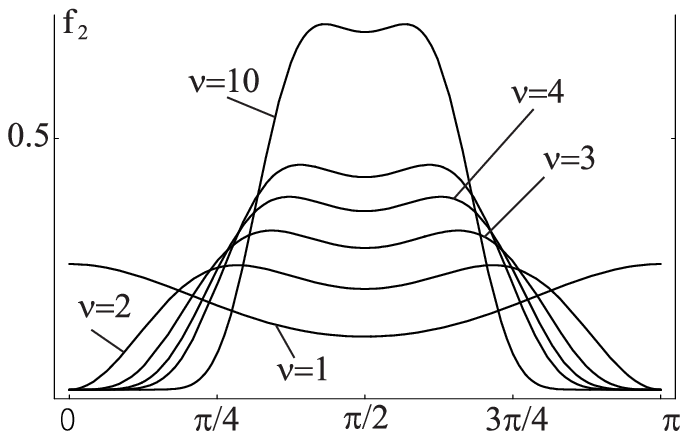}%
\end{center}
%EndExpansion%
%TCIMACRO{\FRAME{ftbphFU}{2.9213in}{1.7314in}{0pt}{\Qcb{The functions
%$f_{k}(\nu,1;\theta),k=0,1,2,3$ at $\nu=1,2,3,4,5,10$.}}{\Qlb{fig6}%
%}{fig_d3_1.eps}{\special{ language "Scientific Word";  type "GRAPHIC";
%width 2.9213in;  height 1.7314in;  depth 0pt;  original-width 2.7916in;
%original-height 1.644in;  cropleft "0";  croptop "1";  cropright "1";
%cropbottom "0";  filename 'fig_d3_1.eps';file-properties "XNPEU";}}}%
%BeginExpansion
\begin{figure}
[ptbh]
\begin{center}
\includegraphics[
height=1.7314in,
width=2.9213in
]%
{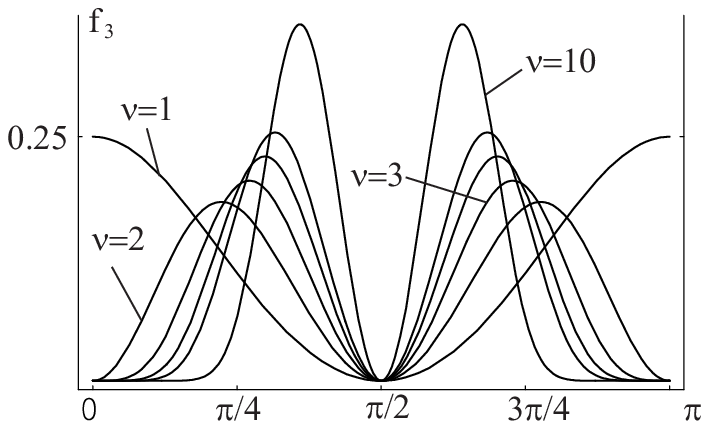}%
\caption{The functions $f_{k}(\nu,1;\theta),k=0,1,2,3$ at $\nu=1,2,3,4,5,10$.}%
\label{fig6}%
\end{center}
\end{figure}
%EndExpansion

Integrating a spectral SR-intensity of $k-$polarization component over all the
directions, one can obtain the so called total spectral SR-intensity of
$k-$polarization component. Let us denote via $\nu_{k}^{\max}=\varphi
_{k}\left(  \beta\right)  $ the harmonic that has a maximal total spectral
SR-intensity of $k-$polarization component. Then, for any given $\nu$ there
exist functions $\beta_{k}(\nu)$ such that
\[
\varphi_{k}\left(  \beta\right)  =\mathrm{const}=\nu\,,\;\beta\in\left[
\beta_{k}(\nu),\beta_{k}(\nu+1)\right]  \,.
\]
In the article \cite{5} the functions $\beta_{k}(\nu)$ were studied in detail.
Results of such an analysis, being compared with the above consideration,
allow us to conclude that the function $\delta_{0}(\nu,\beta)$ is not zero for
$\nu=\nu_{0}^{\max}=\varphi_{0}\left(  \beta\right)  $ . The function
$\delta_{2}(\nu,\beta)$ equals zero for $\nu=\nu_{\sigma}^{\max}%
=\varphi_{\sigma}\left(  \beta\right)  $ that corresponds to $\sigma
$--polarization component of the total spectral SR-intensity.

\textbf{Acknowledgment} The authors (V.G.B., V.B.T., and A.T.J.) thank RFBR
and Minobrazovanie RF for partial support; (D.M.G, and J.A.J) are thankful to
FAPESP for support; (D.M.G) thanks also CNPq for permanent support.

\end{document}